\documentclass[reqno ,12pt]{amsart}
\usepackage[foot]{amsaddr}
\usepackage{graphicx}
\usepackage[usenames,dvipsnames]{xcolor}
\usepackage[paperwidth=8.5in,paperheight=11in,text={6in,9in},left=1.25in,top=1in,headheight=0.25in,headsep=0.4in,footskip=0.4in]{geometry}
\usepackage{subfigure}
\usepackage{lineno}
\usepackage{url}
\usepackage[natbibapa]{apacite}

\definecolor{linenocolor}{gray}{0.6}

\usepackage{fix-cm}

\begin{document}

\title[Adoption as a Social Marker]{\large Adoption as a Social Marker: \\ Innovation Diffusion with Outgroup Aversion}
\author[Smaldino, Janssen, Hillis, \& Bednar]{Paul E.~Smaldino$^{1,*}$ \and Marco A.~Janssen$^2$ \and Vicken Hillis$^3$ \and Jenna Bednar$^4$}
\address{$^1$Cognitive and Information Sciences, University of California, Merced}
\address{$^2$School of Sustainability, Arizona State University, Tempe}
\address{$^3$Department of Environmental Science and Policy, University of California, Davis}
\address{$^4$Department of Political Science, University of Michigan, Ann Arbor}
\email{paul.smaldino@gmail.com}

\maketitle

{\vspace{-6pt}\footnotesize\begin{center}\today\end{center}\vspace{24pt}}


\noindent\textsc{Abstract.} 
Social identities are among the key factors driving behavior in complex societies. Signals of social identity are known to influence individual behaviors in the adoption of innovations. Yet the population-level consequences of identity signaling on the diffusion of innovations are largely unknown. Here we use both analytical and agent-based modeling to consider the spread of a beneficial innovation in a structured population in which there exist two groups who are averse to being mistaken for each other. We investigate the dynamics of adoption and consider the role of structural factors such as demographic skew and communication scale on population-level outcomes. We find that outgroup aversion can lead to adoption being delayed or suppressed in one group, and that population-wide underadoption is common. Comparing the two models, we find that differential adoption can arise due to structural constraints on information flow even in the absence of intrinsic between-group differences in adoption rates. Further, we find that patterns of polarization in adoption at both local and global scales depend on the details of demographic organization and the scale of communication. This research has particular relevance to widely beneficial but identity-relevant products and behaviors, such as green technologies, where overall levels of adoption determine the positive benefits that accrue to society at large.

\vspace{12pt}


\vspace{24pt}
\newpage

\section{Introduction}

The adoption of new technologies, products, and behaviors can be influenced by identity signaling. Consider, as an example, the adoption of hybrid cars. Owners of a Toyota Prius are willing to pay a few thousand dollars to signal their green intentions \citep{sexton2014}. On the other hand anti-environmentalists spend thousands of dollars to modify their vehicle's emissions systems to deliberately spew sooty diesel exhaust on electric cars, bicyclists and pedestrians \citep{weigel2014}. In 2014, Cadillac and Ford capitalized on differences {\em among} proponents of hybrid cars, producing advertisements for their plug-in hybrids that appealed to contrasting values among car owners for products with similar functional qualities \citep{bradford2014}. Recent experimental work has additionally highlighted that both adoption and disadoption of products and other consumptive behaviors are influenced by the perceived social identities of previous adopters \citep{berger2007, berger2008, brooks2015, morvinski2014}. It is therefore clear that social identity should not be ignored when considering how products diffuse in a population. Nevertheless, the role of identity has been largely absent in models of product diffusion \citep{peres2010}. 

In this paper, we present formal models of product adoption under the influence of identity signaling and explore the population-level consequences of identity on diffusion dynamics\footnote{We use the terms ``product" and ``innovation" interchangeably, as our models apply equally to innovations, non-innovative products, and any other behavior that can be adopted via social influence.}. 
We examine product adoption as a form of social marking, and how the emergence of a product's role in identity signaling influences the diffusion of innovations.  In particular, we investigate the effect of outgroup aversion on adoption, taking into account the spatial structure of a population and varying scales of communication across distances. In the interest of clarity, we first discuss a simple analytical model that relies on coupled differential equations. This model was first introduced by Bakshi et al. (2013); we adapt it for our purposes.  We then introduce an agent-based model with explicit spatial structure.  A model capable of capturing who interacts with whom, as agent-based models do, is critical for understanding social behavior because the structure of interactions shapes the temporal and categorical dynamics of social behavior \citep{boccaletti2006, durrett1994, newman2003}, including how innovations spread \citep{burt1980, abrahamson1997, davis1997, delre2010, watts2007}. We demonstrate that 1) differential rates of adoption typically attributed to intrinsic between-group differences in rates of adoption can arise simply due to structural constraints on information flow, and 2) that equilibrium levels of adoption, as well as global and local polarization in adoption, depend heavily on both demographic organization and the scale of communication.

\subsection{Identity Signaling and Product Adoption}
Identity signaling is the broadcasting of one's membership in some group or collective, with receivers including both fellow members and non-members. Identity signaling is important when multiple groups exist in a population and individuals benefit by being correctly identified with their group, as well as by not being mistaken for a member of another group. The anthropological literature has long characterized arbitrary social markers as critical for coordination when correlations exist between group identity and behavioral norms \citep{barth1969, mcelreath2003, moffett2013human, wimmer2013}. In complex societies such as ours, interactions delineated by social identities are ubiquitous \citep{smaldino2017}, and social markers are crucial coordination tools. Although social identities are multidimensional and context dependent \citep{ashmore2004, roccas2002}, in any particular context humans appear to have strong instincts to identify with a group, even if the distinguishing factor between ingroup and outgroup is arbitrary \citep{tajfel1971}. 

Although individuals often adopt a product for its intrinsic value, product adoption can also be used to distinguish oneself from other groups \citep{berger2008}. When adoption can be observed by others, products may become social markers and serve the role of helping individuals to find appropriate social partners for interactions \citep{mcelreath2003, smaldino2017}. For this reason, an individual may be less likely to adopt an innovation if it is more strongly associated with an outgroup than with their ingroup or with no group at all. Indeed, recent marketing studies suggest that the identity signaling component of product adoption comes into play even when the outgroup does not evoke negative affect, as long as it is seen as dissimilar. For example, in one study, Stanford undergraduates evaluated an mp3 player substantially less favorably after being told that it was popular with ``business executives" (whom they rated as dissimilar but not disliked) compared with when they were told it was popular with ``individuals" (Berger \& Heath 2007, study 4). Conversely, \cite{morvinski2014} found that information about a large number of previous adopters positively influenced adoption only if those previous adopters were described as similar to the potential adopters.  In another study, after adopting a charity-affiliated bracelet, students from a ``jock" dorm dis-adopted (i.e., stopped wearing the bracelet) when the bracelet started appearing on the wrists of students from the nearby ``nerdy" dorm, whom the jocks did not dislike but also did not want to be mistaken for (Berger \& Heath 2008, study 2). This latter study also serves to illustrate that products need not be intrinsically associated with particular social groups, but can emerge as social markers through preferential adoption by one group, for reasons that may be initially arbitrary. 

\subsection{Population Structure and Population Dynamics}
The sociological literature has shown that even small differences in opinions, preferences, or product adoption can cascade, through the processes of homophily and social influence, into highly polarized social networks \citep{axelrod1997, carley1991, flache2011, mark1998}. A great deal of research has shown that the structure of those networks affects exactly how individuals influence, and are influenced by, one another  \citep{banerjee2013, burt1980, abrahamson1997, delre2010, flache2011, funk2009, strogatz2001, watts2007}. As a striking example, conformity across a broad spectrum of lifestyle preferences is seen within clusters of Americans identified only by political affiliation, such that arbitrary traits become markers of identity, as with ``latte-drinking liberals" and ``gun-toting republicans" \citep{dellaposta2015}. Although many factors lead to the adoption of behaviors or technologies, social influence through no other factor than network proximity can be such a powerful force that some have argued it must be explicitly discounted before other cultural or developmental explanations may be considered \citep{dellaposta2015, dow1982, mcpherson2004}.

As a description of the emergence of an initially neutral product becoming strongly associated with a particular social identity, consider another example, this time from the beer industry. In the 1980s, Pabst Blue Ribbon (PBR) was a relatively unpopular beer in the United States, loosely associated with the white working class. Then, in the late 1990s, bars in Brooklyn and Lower Manhattan began offering PBR drink specials, attracted by its low price and relative obscurity. The brand gradually became associated with urban hipsters, spread to hipster enclaves in Portland and Los Angeles, and has since found widespread adoption in hip youth culture. Without any direct advertising, adoption as a social marker cascaded. Sold in 1985 for \$63 million (\$141 million in 2015 dollars), PBR was recently sold again in 2014 for an estimated \$700 million \citep{gelles2014}. Meanwhile, its identity as a ``hipster beer" may have turned at least some people away. As of this writing, the top definition for PBR on urbandictionary.com includes the remark, ``Pabst Blue Ribbon is a lot like the band Bright Eyes. Hipsters love it, but everyone else thinks it's liquid shit."

A standard model of product adoption invokes status: a social marker describes a demarcation between social elites, who innovate in order to distinguish themselves, and the lower classes, who imitate the elites. This process of chase and flight drives new cycles of fashions as the elites continuously attempt to distance themselves from the lower classes \citep{simmel1957}. However, many cases of adoption or abandonment of products are not so easily characterized in terms of status. Indeed, innovations and fashions often originate from among the lower or middle classes \citep{berger2008}. Chase and flight dynamics are undoubtedly important in many settings, including some that don't involve groups with differing levels of socio-economic status \citep{bakshi2013}. However, products are commonly used as social markers facilitating preferential interaction with ingroup members on the part of {\em all} groups. Indeed, a large anthropological literature has demonstrated that, regardless of status, individuals are often better off when they can easily distinguish group members from non-members \citep{barth1969, mcelreath2003, moffett2013human, wimmer2013}.  It is therefore important to examine diffusion dynamics for scenarios involving general tendencies for outgroup aversion, as suggested in an influential review by \cite{peres2010}. Such an analysis has hitherto been missing from the literature on product adoption.
 
There are many important factors to consider in the dynamics of innovation adoption, and our aim here is not to cover all of them with the veneer of identity signaling. In particular, we will not address multi-brand competition, in which companies selling similar products compete for customers. This type of competition has only recently been the subject of formal modeling, always with the assumption that customers are identical and brands equivalent \citep{libai2009, laciana2014a, laciana2014b}. Such competition entails rich dynamics, influenced by a host of factors such as the timing of product release, the network position of early adopters, and the strength of cross-brand influence \citep{libai2009, libai2013}. Because the interplay between adoption and identity signaling has not been previously modeled, we believe it appropriate to limit our study here to the one-product, one-brand case so as to establish a baseline. Later in this paper, we will discuss multi-brand competition in light of our results.   

\subsection{Modeling Approach}
We wish to consider how patterns of adoption are influenced by {\em outgroup aversion}, defined as a desire to not engage in activities associated with the outgroup;  {\em demographic organization}, including network autocorrelation; and the {\em scale of communication}, contrasting local, short-range observations with long-range observations of distant actors. We also examine how these three factors influence patterns of local and global polarization, or the extent to which adoption is skewed towards one group or another, suggesting demographic segregation.  In the following sections, we present two models of innovation adoption as a social marker. The first is an analytical model that takes as its starting point the well-known model of \cite{bass1969}. This was the first model to provide a generative explanation of why many diffusion processes generate S-shaped adoption curves. We extend this model to allow for two groups who are averse to products associated with the other group. Such a model is necessarily limited, however, because it assumes perfect population mixing and therefore cannot account for the influences of structural aspects of demography or communication. Focusing on structural influences can provide important insights, because micro-level mechanisms frequently aggregate into population-level consequences in non-obvious ways \citep{davis2006}.  To investigate these influences, therefore, we introduce a spatially-structured agent-based model. 

Our agent-based model allows for quite a bit more complexity than is afforded by the analytical model, and thus a {\em direct} comparison is not our intention. However, formalizing the verbal description in two distinct ways provides several benefits. First, while both models illustrate that when products become social markers, outgroup aversion can lead to population-wide underadoption as well as delayed or suppressed adoption among members of one group, each model highlights a distinct causal mechanism. Second, the analytical model provides a formal link between prior work on the diffusion of innovations, while the agent-based model that allows us to more fully explore our questions of interest. Third, there are often many possible formal instantiations of a complex social system. We show that two different formalizations yield compatible results, thus providing the beginnings of a triangulation on a robust theory of social identity and product adoption. Finally, we also connect our model with the literature on opinion polarization, examining the factors that lead to different kinds of polarization in product adoption.

\section{Analytical Model}
The canonical model of product diffusion was introduced by \cite{bass1969}, and provided a concise explanation of the sigmoidal patterns of diffusion observed by \cite{rogers1962} and others \citep{bass2004, barnett2011}.  \cite{bakshi2013}  recently extended Bass's model to account for product adoption by two interacting groups\footnote{Bakshi et al. refer to these as ``segments" of the population.}. We adopt their formalism here. 
For each group, the probability of adoption is influenced by three factors: a background rate of spontaneous adoption, social influence from one's group members, and social influence from members of the outgroup. The model takes the form of coupled differential equations
\begin{align}
\frac{dN_1}{dt} &= (a_1 + b_1 N_1 + c_1 N_2) (m_1 - N_1)  \\ 
\frac{dN_2}{dt} &= (a_2 + b_2 N_2 + c_2 N_1)(m_2 - N_2),
\end{align}
where for each group $i$, $a_i > 0$, $b_i > 0$, $m_i > 0$, and $0 \leq N_i \leq m_i$. $N_i$ is the current number of adopters at time t, out of a potential market of size $m_i$. The remaining terms are the coefficient for innovation, $a_i$, the coefficient for within-group imitation, $b_i$, and the coefficient for cross-group imitation, $c_i$. If both $c$ terms are zero, the model reduces to two independent instantiations of the Bass (1969) model. \cite{bakshi2013} took their inspiration from Simmel's (1904/1957)\nocite{simmel1957} chase-and-flight theory, and restricted their analyses to cases where cross-group influence was positive for one group and negative for the other. Here, we are interested in the scenario where both groups want to avoid being associated with the opposite group, and so will reject or even disadopt products that are popular with the opposite group \citep{barth1969, berger2008, wimmer2013}. We therefore focus on the case where both $c_1$ and $c_2$ are less than zero.    

\begin{figure}[tp]
\centerline{\includegraphics[width=.99\textwidth]{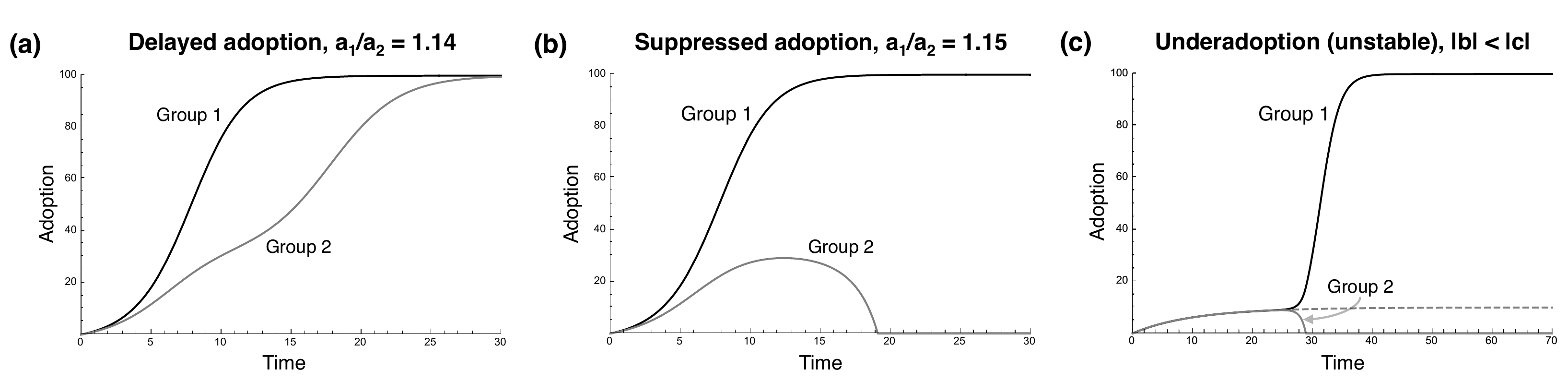}}
\caption{Numerical simulation of the analytical model. For all simulations, $a_2 = .01$, $b = .006$, $m = 100$. When ingroup attraction is stronger than outgroup aversion and one group has a higher innovation rate, the group with the lower innovation rate can have delayed (a) or suppressed (b) adoption. For these simulations, $c = -.002$. When outgroup aversion is stronger than ingroup attraction, an unstable equilibrium of mutual partial adoption can give way to complete adoption by one group and zero adoption by the other (c). For this simulation, $a_1 = a2 = .01$, $c = -.007$.}
\label{bassdyn}
\end{figure}

A closed-form solution for this system of equations is not analytically tractable because of the coupled dependencies. Moreover, we are heavily interested in demographic and communication effects related to population structure, so we forego a complete specification of the analytical model system\footnote{We provide, as a supplement, a Mathematica notebook for reproducing and altering the plots shown in Figure \ref{bassdyn} to enable a more thorough exploration of the analytical model by the curious reader.}. However, some numerical analysis is quite revealing. Our primary interests are in the dynamics of adoption when products are social markers in the case of mutual repulsion, and not in the consequences of major between-group differences in within and between-group imitation or market size (though such consequences are surely worthy of future investigation). As such, we focus on scenarios where the two groups are largely identical in their propensity for within-group imitation, cross-group imitation, and market size, so that $b_1 = b_2 = b$,  $c_1 = c_2 = c$,  $m_1 = m_2 = m$. 

We anticipate that three categories of outcomes might emerge from the interaction of outgroup aversion, social influence, and innovation rate. When outgroup aversion is particularly strong it might overcome the force of ingroup imitation entirely and result in underadoption in both groups. When ingroup imitation outweighs outgroup aversion, we expect adoption to occur, and to be quicker in the group with a higher rate of innovation. To explore these dynamics in more detail and their resultant equilibria, we vary the strengths of outgroup aversion, ingroup imitation, and innovation rates, as described below.
 {\em Delayed adoption} can occur if outgroup aversion is weaker than ingroup imitation, $|c| < |b|$, and the difference between the rates of innovation in each group is small. If innovation rates are equal, $a_1 = a_2$, then adoption will increase at the same rate in both groups, and since ingroup attraction outweighs outgroup repulsion, both groups will always achieve full adoption at the same time. If, however, the innovation rate is slightly higher in one group, then that group will adopt at an initially faster rate. Even if this advantage in innovation rate in the first group is small, adoption among members of the second group can be substantially delayed (Figure \ref{bassdyn}a). If this advantage is only slightly larger, however, adoption in the second group can be completely {\em suppressed}. In our example, shown in Figure \ref{bassdyn}b, an advantage of only 15\%  in innovation rate leads to enough of a disparity in adoption rates between groups that outgroup aversion comes to completely suppress long-term adoption in the group slower to innovate. Finally, consider if outgroup aversion is stronger than ingroup attraction, $|c| > |b|$. Whatever group is slower to adopt initially will eventually be entirely suppressed. If both groups innovate at exactly the same rate, a state of mutual {\em underadoption} can occur, in which neither group adopts at saturation levels. However, such a state is unstable, and even the slightest perturbation will lead to full adoption by one group and zero adoption by the other (Figure \ref{bassdyn}c). 

This analysis indicates how the Bass (1969) model can be extended to include mutually aversive social groups, and that outgroup aversion can lead to delayed adoption or complete suppression of adoption by one group. Mutual underadoption is possible but unstable. Thus, we may begin to investigate the appearance of these trends in the dynamics of real world adoption. 

Our aim is to advance the model from the stylized analytical model in a direction that better matches reality. The analytical model assumes a well-mixed population and random interactions among all individuals regardless of group membership. These assumptions are useful as a first approximation but ultimately unsatisfying. Members of different social groups do interact with one another, but individuals often preferentially interact with members of their own social groups \citep{lazarsfeld1954, mcpherson2001}. Moreover, communication occurs both locally and occasionally over long distances, and rarely resembles the well-mixed population implicit in the Bass model. For these reasons, we turn to an agent-based model in order to investigate the effects of structure in demography and communication. 

Agent-based models involve explicit instantiation of individuals, and through computer simulation can explore consequences of heterogeneous individuals interacting in structured populations in ways difficult or impossible for purely mathematical approaches\footnote{Even when mathematical approximations are possible, they are imprecise and can miss important results only available to simulation approaches, as in \cite{deaguiar2004}.}  \citep{epstein2006, peres2010, rand2011, smaldino2015}. We may then see if and when the outcomes generated by the analytical model occur in the agent-based model, and perform more nuanced analyses of the dynamics of diffusion.

\section{Agent-Based Model}

We designed our model not to represent any {\em specific} demographic system, but rather as an abstraction representing {\em general} properties common to many systems in which individuals maintain social networks on which many but not all interactions take place. The population is structured into discrete patches upon which agents live and interact. Patches can represent geographic localities, but need not; they simply connote any highly clustered interaction network. Inter-patch interactions may also occur, representing long-distance effects such as travel, telephone, and social media. A product will be introduced which has some intrinsic appeal. As in the analytical model, the probability of adoption increases with multiple exposures, reflecting sociological research on complex contagion \citep{centola2007, centola2010} as well as marketing research on network externalities, which can increase the utility of adoption as more people adopt \citep{peres2010}. However, if an individual's experience with the product is such that its adoption is preferentially associated with the outgroup, adoption becomes {\em less} likely. In our analysis we will explore the effect of several parameters, including the importance of identity signaling, the extent of homophily in demographic organization, and the influence of long-range interactions.

\begin{figure}[tp]
\centerline{\includegraphics[width=.6\textwidth]{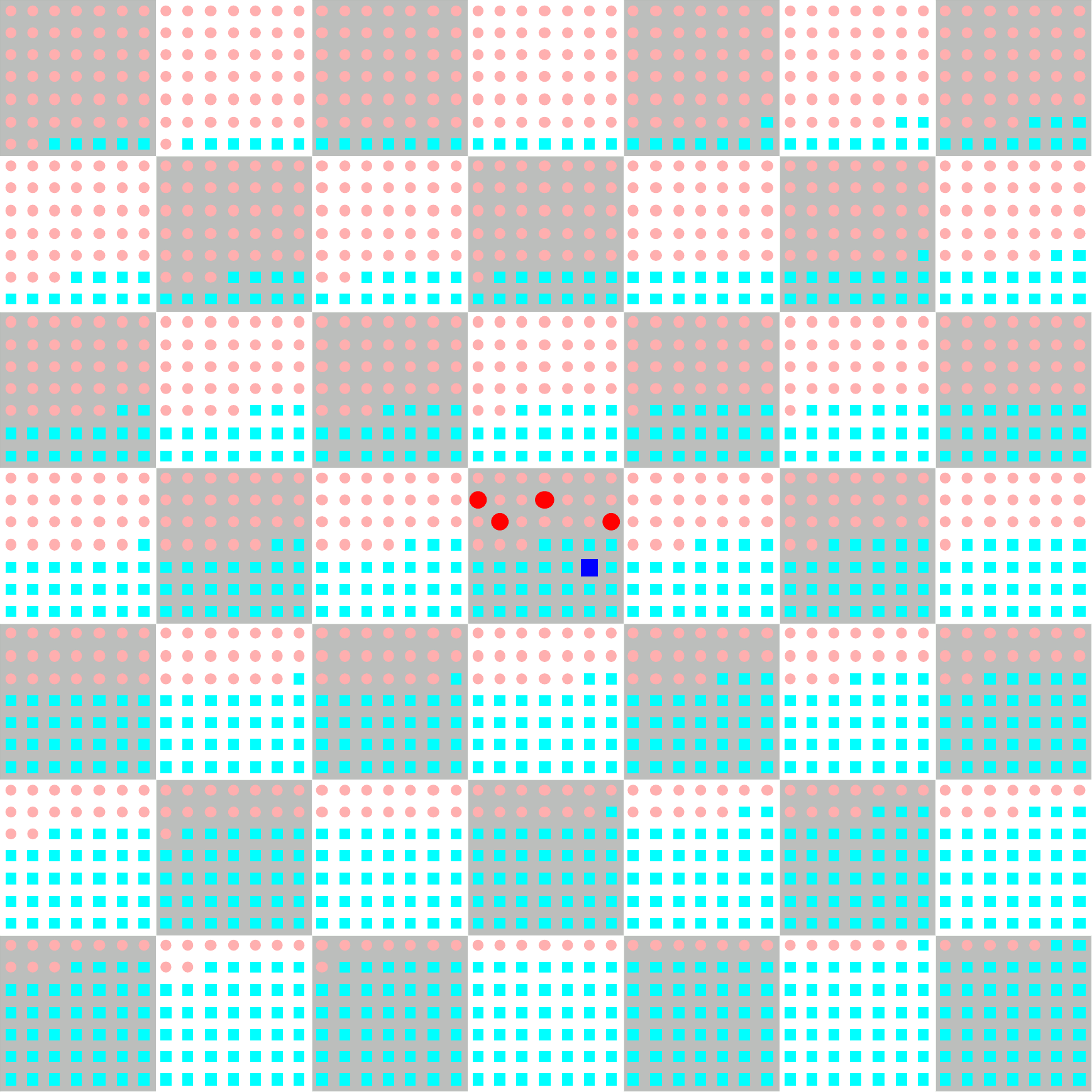}}
\caption{Schematic visualization of the model world. Alternating light and dark squares are patches. Pink circles are members of group $A$ and blue squares are members of group $B$. Five early adopters in the center patch (\#25), four from group $A$ and one from group $B$, are indicated by the darker red and blue coloring and increased size. Here $M = N = 49$, and $Q = 0.9$. }
\label{modellayout}
\end{figure}

More formally, the population is distributed across $M$ patches, which are situated along a line so that each patch has two neighboring patches with the exception of patches 1 and $M$ (the first and last patches), which have only one neighbor.  While our line of patches is not as common a way of modeling spatial structure as others, such as Moore neighborhoods, we settled on this arrangement because it facilitates the examination of short and long range interactions among individuals with varying identities in the most minimal way\footnote{The line is the simplest organizational framework with which to study structured interactions. It also allowed for an easily implementable and interpretable spatial correlation between neighboring patches, which would be more complicated in network structures in which patches could share more than one neighbor in common, such as a square lattice or small world network. That said, our model is easily extendable to those and other network structures.}.  We visualized this layout by placing the patches on a square grid, such that the rightmost patch in one row neighbors the leftmost patch in the row just below it (Figure \ref{modellayout}). Each patch contains $N$ agents. Each agent has one of $S$ social identities ($S = \{A, B\}$). To avoid global majority/minority effects, we assumed equal overall numbers of individuals holding each group identity, though we allowed the distribution of identities to vary within each patch. Because similar individuals tend to assort \citep{dow1982, mcpherson2001}, we generally assumed that, firstly, each patch would tend to be dominated by individuals belonging to one group or the other, and secondly, that the demographic skew of patches (i.e., the proportion of agents having a particular social identity) would tend to be spatially correlated -- that the identity distribution within each patch more closely resembles the distribution in neighboring patches than in distant ones\footnote{In assuming the spatial correlation between patches we have in mind the smooth transitions that one experiences in moving from one media market to another and not the sharp distinctions in identities that one experiences when crossing the tracks or suburban boundaries, as in a \cite{tiebout1956} world.}.   
The algorithm for assigning agents to patches is given
below. At time $t = 0$, an innovation is introduced into the population, which can be adopted.  We assume that all agents share identical intrinsic valuation for the product and are equally capable of adopting it. In a single patch, $n_0 < N$ agents are chosen as ``early adopters," who are initialized as already having adopted the innovation, and are randomly drawn from the same patch without heed to social identity. Note that we use the word ``patch" to denote agents' spatial location. All uses of the word ``group" refer to agents' social identity, not their location.

Agents learn about the innovation through interactions with other agents. At each time step, each agent makes an observation with probability $\mu$, the interaction rate. If an observation occurs, the focal agent observes a sample of $m$ other agents. These observed agents are sampled one at a time. With probability $f$, an observed agent is sampled at random from the focal agent's own patch. Otherwise, the observed agent is drawn at random from an external patch. The parameter $f$ allows us to continuously vary the clustering of the population, with a well-mixed population at one extreme and an isolated islands model at the other. When observations involve an external patch, that patch may be restricted to spatial neighbors of the focal agent's patch ({\em local} external observation), or be randomly drawn from the entire population of patches ({\em long-range} external observation). Neighboring patches have correlated demographic properties; that is, they have similar proportions of the two marked groups. Long-range observations result in a more well-mixed population structure in which individuals are influenced by others who might reside in environments that are, at a local level, demographically quite different\footnote{In reality, even long-range observations will likely be skewed by homophily. Our formalization allows us to control the rate of homophilic interactions more precisely.} \citep{flache2011}. 
Observations allow agents to become aware of innovations, and to make assessments of the innovation's prevalence among both in- and out-group individuals\footnote{Product information flow in our model involves only direct social communication, and for simplicity excludes mass media influences. A modification in which agents received additional information from third-party sources, perhaps represented by ``media agents," could be added for future analyses.}. The order in which agents make observations is randomized at each time step. Once an agent has observed $m$ other agents, she makes a decision about whether to adopt the innovation.

\subsection{Social Influence and Innovation Adoption}
Following an observation, the agent decides whether or not to adopt the innovation. This decision is made even if the agent has previously adopted the innovation; in this case, a decision not to adopt implies a disadoption. Two factors contribute to the consideration of adoption: (1) a generalized frequency-dependent bias, whereby the more popular a product is in the population, the more likely an agent is to 
adopt it, and (2) an aversion bias, whereby agents prefer not to adopt an innovation adopted by outgroup members. By including a generalized frequency-dependent bias instead of restricting positive influence to one's ingroup, we are able to analyze the contrasting influences of positive frequency dependence and outgroup aversion in a more realistic manner than was possible in the analytical model.   

First, consider the frequency dependent influence, $F$. An agent is more likely to adopt a more common product. We follow Granovetter (1978) in assuming that the likelihood of adoption is influenced by positive feedback -- the more people do something, the more people will be willing to join in. This influence is probabilistic; for example, Berger and Heath (2007) found that, regardless of whether a product was identity relevant, some small proportion of people (at least 14\% in their study 3) still chose a minority product over a more popular one. Nevertheless, their findings support the claim that, all things being equal, people tend to prefer more popular items, fitting the literature on positive biases in social transmission of ideas and behaviors \citep{bikhchandani1998}. In many cases, social forces can also lead to the preferential adoption of already popular items. For example, \cite{salganik2006} used an artificial online music market to show that a product's adoption was largely determined by its early popularity, independent of inherent quality. Our assumption of positive frequency dependence is also congruent with research on complex contagion, which indicates that the adoption of social behavior often requires multiple sources of influence \citep{centola2007}, as well as with research on network externalities, in which the utility of adoption increases as more consumers adopt the product \citep{peres2010}. For example, the utility of purchasing a plug-in electric car increases as more people own them, because charging stations will become more prevalent and mechanics will be encouraged to gain expertise in their maintenance.  

The positive frequency-dependent bias is tempered by the effects of outgroup aversion, the inverse of which is denoted by $V$. An agent is more likely to adopt an innovation that is more closely associated with individuals of the same social identity, and less likely to adopt an innovation that is closely linked to agents of another social identity. There is considerable evidence that people are more like to adopt a product when it is preferentially associated with their ingroup \citep{morvinski2014}, and that people who initially adopt a behavior or product will abandon it if it becomes associated with outgroup individuals \citep{berger2007, berger2008}. \cite{bakshi2013} observe that ``Repulsion [from the outgroup] results in flight from potential adopters, but does it also lead prior adopters to flee and disadopt? It very often does, the one exception being situations with very high switching costs." (p. 3).  In agreement with \cite{bakshi2013}, our model is most relevant to products with low to moderate switching costs. We note, however, that even expensive items may have low switching costs. For example, in the U.S. it is not uncommon to purchase a new car every few years.

We assume that frequency dependence and outgroup aversion interact multiplicatively, so that the probability of adoption is given by 
\begin{equation}
\Pr(\text{adopt}) = FV. 
\end{equation}
The frequency-dependent component is given by
\begin{equation}
F = x^\lambda,
\end{equation}
where $x$ is the proportion of observed agents who have adopted the innovation, and $\lambda \in (0, 1)$ is a control parameter representing the strength of the frequency dependence. Many models of innovation diffusion formalize positive bias by a constant parameter times the number of adopters encountered, as we did in our analytical model. For our agent-based model, we wanted to be able to control the total number of agents observed, corresponding to an overall rate of contact, without changing the overall likelihood of adoption. Indeed, someone whose interactions are restricted to a small group of friends, all of whom have adopted, should be more likely to adopt than someone who has encountered many more people, only a fraction of whom have adopted. The parameter $\lambda$ allows us to maintain this relationship between observations and adoptions and still vary the strength of the frequency dependence. When $\lambda$ is closer to zero, the effect of direct social influence is small, so that innovations are likely to be adopted even at fairly low frequencies. When $\lambda$ is closer to one, the probability of adoption scales linearly with the observed frequency of adopters\footnote{Mathematically, it is of course possible for $\lambda$ to be greater than one, further decreasing the probability of adoption when the product is rare. We restrict our analysis to the range (0, 1), because diffusion tends to fail for values of $\lambda$ near or greater than one.}. Note also that this formulation conveniently combines socially influenced adoption and spontaneous innovation, which are separated in the Bass model, into a single term. 

\begin{figure}[tp]
\centerline{\includegraphics[width=.8\textwidth]{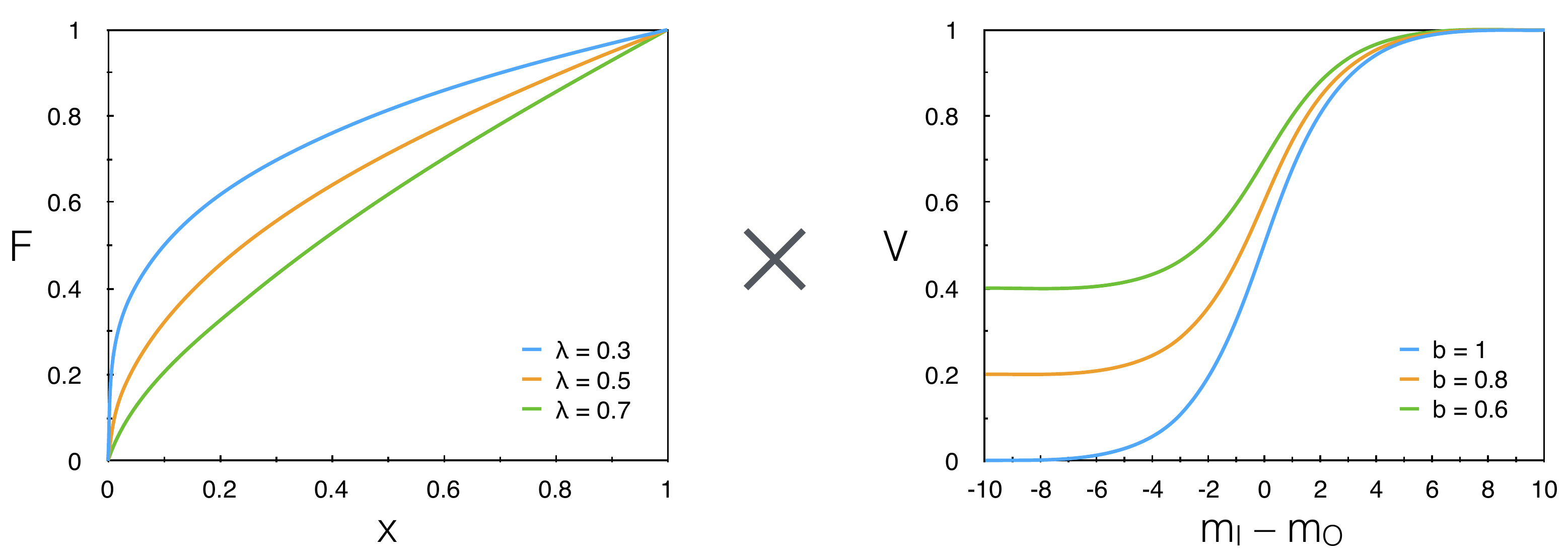}}
\caption{The probability of adoption is the product of $F$ and $V$.}
\label{adoptfunc}
\end{figure}

The influence of ingroup-outgroup bias is given by
\begin{equation}
V = 1 - b + \frac{b}{1 + \exp[-(m_I - m_O )]},
\end{equation}				
where $m_I$ and $m_O$ are the number of observed agents who have adopted the innovation and are either in the ingroup or the outgroup, respectively, and $b \in [0,1]$ is the strength of outgroup aversion, such that the lower bound of $V$ is $1 - b$. The functions $F$ and $V$ are visualized in Figure \ref{adoptfunc}.

\subsection{Initialization}
The population consists of $M$ patches, each with $N$ agents, for a total of $MN$ agents. We assume that there are equal numbers of each group's members in the overall population. However, individual patches can vary in the proportions of members of each social group. The demographic skew, $Q$, is the maximum proportion of group $A$ among the patches, so that the patch with the lowest frequency of group $A$ members has frequency $1 - Q$. For each patch $j$, such that $j \in \{1,...,M\}$, the frequency of group $A$ members $q_j$ is given by 
\begin{equation}
q_j = \left(1 - \frac{j - 1}{M - 1} \right) Q + \left( \frac{j - 1}{M - 1} \right) (1 - Q),
\end{equation}
which simplifies to 
\begin{equation}
q_j = \frac{j - 1}{M - 1} + \left(1 - \frac{2(j - 1)}{M - 1} \right) Q.	
\end{equation}	
Patch $j$ therefore contains $q_j N$ members of group $A$ and $(1 - q_j)N$ members of group $B$, both rounded to their nearest integer values. Note that when $Q = 0.5$, each patch has an equal number of members from each group. $Q = 1$ is the maximum skew, where each group has a single patch entirely devoid of outgroup individuals. For our initial simulations, we used $Q = 0.9$, and $M = N = 49$ (see Figure \ref{modellayout}). For most simulations, five early adopters were seeded in the most centrally located and heterogeneous patch (patch 25, the center patch), though we also experimented with seeding early adopters in a more far-flung, more homogeneous patch (patch 1, the leftmost patch). All parameters and values used for our simulations are summarized in Table \ref{table_parameters}. 

\begin{table}
\small
\begin{center}
\begin{tabular}{|l|l|l|l|}
\hline
{\bf Parameter} 	&	{\bf Definition}	&	{\bf Default Value} & {\bf Sensitivity}\\
\hline
$n_0$ 		&	number of early adopters						& 5 		&	\\
$\mu$ 		&	interaction rate									& 0.05 	&	\\
$m$ 			&	number of agents observed					& 30 		&	\\
$f$ 			&	probability of within-patch observation		& 0.7 		&	$[0.1, 0.9]$\\
$\lambda$ 	&	frequency dependence						& 0.3 		&	$[0.1, 0.9]$\\
$b$ 			&	outgroup aversion								& 1 		&	$[0,1]$\\
$Q$ 			&	demographic skew								& 0.9 		&	$\{0.5, 0.9\}$\\
Start patch	&	location of early adopters						& 25 		&	$\{1, 25\}$\\
External observation	&	location of observed extra-patch agents		& 0.7 		&	\{local, random\}\\
\hline
\end{tabular}
\end{center}
\caption{Model parameters and values. Sensitivity is the set or range of values tested.}
\label{table_parameters}
\end{table}%

\subsection{Metrics of Adoption and Polarization}
Each model instantiation was run for 2,000 time steps, which was always long enough to reach an equilibrium in which adoption levels remained consistent. Inspection of sample simulations run for over 20,000 time steps indicates that such equilibria are dynamically stable. Aggregate data are averages from 100 runs for each set of parameter values. Our results are summarized using the total {\em adoption} level as well as {\em local} and {\em global polarization} metrics: 	
\begin{itemize}
\item {\em Adoption} is the fraction of the total population that has adopted the innovation. Patch-level adoption levels were also recorded for each of the two social identity groups. 
\item {\em Global polarization} is a measure of the extent to which, at the level of the population, adoption is skewed toward one social identity group over the other. This is simply $|n_1 - n_2|/(n_1 + n_2)$, where $n_1$ and $n_2$ are the number of agents in each group, respectively, who have adopted the innovation. This will be zero when each group has an equal number of adopters and one when all the adopters belong to one group. 
\item {\em Local polarization} is the average patch-level polarization. This is equal to $\sum_j |n_{1j}/N_{1j} - n_{2j}/N_{2j})|$, where $n_{1j}$ and $n_{2j}$ are the number of group $A$ or group $B$ adopters in patch $j$, and $N_{1j}$ and $N_{2j}$ are the total number of group $A$ and group $B$ individuals. This formulation allows us to control for inequalities in the number of agents from each group in each patch, so that a patch where everyone has adopted will have zero polarization regardless of the number of agents from each group. As a convenience, we let $0/0 = 0$. 
\end{itemize}

Our agent-based model was implemented in both Java, using the MASON library \citep{luke2005}, and NetLogo \citep{wilensky1999}, each coded independently by different authors (PS and MJ, respectively) to ensure that results were not due to programming error. The results reported are performed with the Java version and the model code is made freely available on openabm.org\footnote{Model code will be made public when paper is accepted. For now, it is available at: http://smaldino.com/wp/wp-content/uploads/2015/05/AdoptionAsSocialMarker.zip}.  Unless otherwise indicated, all results presented utilize the default parameter values shown in Table 1. 

\section{Simulation Results}

In our analytical model, we found that delayed adoption, suppressed adoption, and mutual underadoption could all occur as a result of differential innovation rates between groups, depending on the strength of outgroup aversion. An important finding of our agent-based investigation is that all three of these outcomes occurred, but with somewhat different characteristics and through different causal mechanisms. In the analytical model, differences in innovation rate corresponded to persistent differences in the spontaneous adoption of the product by members of each group. The empirical meaning of this parameter is unclear, however, unless one can definitively state that members of one group are more or less innovative (though differences in access to the product or inherent wealth could provide a means for such an outcome). In our agent-based model, all adoption apart from a few early adopters is due to direct social influence. Thus, the agent based model demonstrates that differences in group-level adoption outcomes can arise even when there are no differences in innovation rate. Rather, these differences emerge from structural differences in how individuals interact with one another.  

In our agent-based model, the structure of communication is determined by two parameters: the tendency for observations to occur within one's patch, $f$, and whether extra-patch observations are made locally (in a neighboring patch) or randomly across the entire population, with the latter case representing long-range communication or high levels of social mixing. 

\begin{figure}[tp]
\centerline{\includegraphics[width=.95\textwidth]{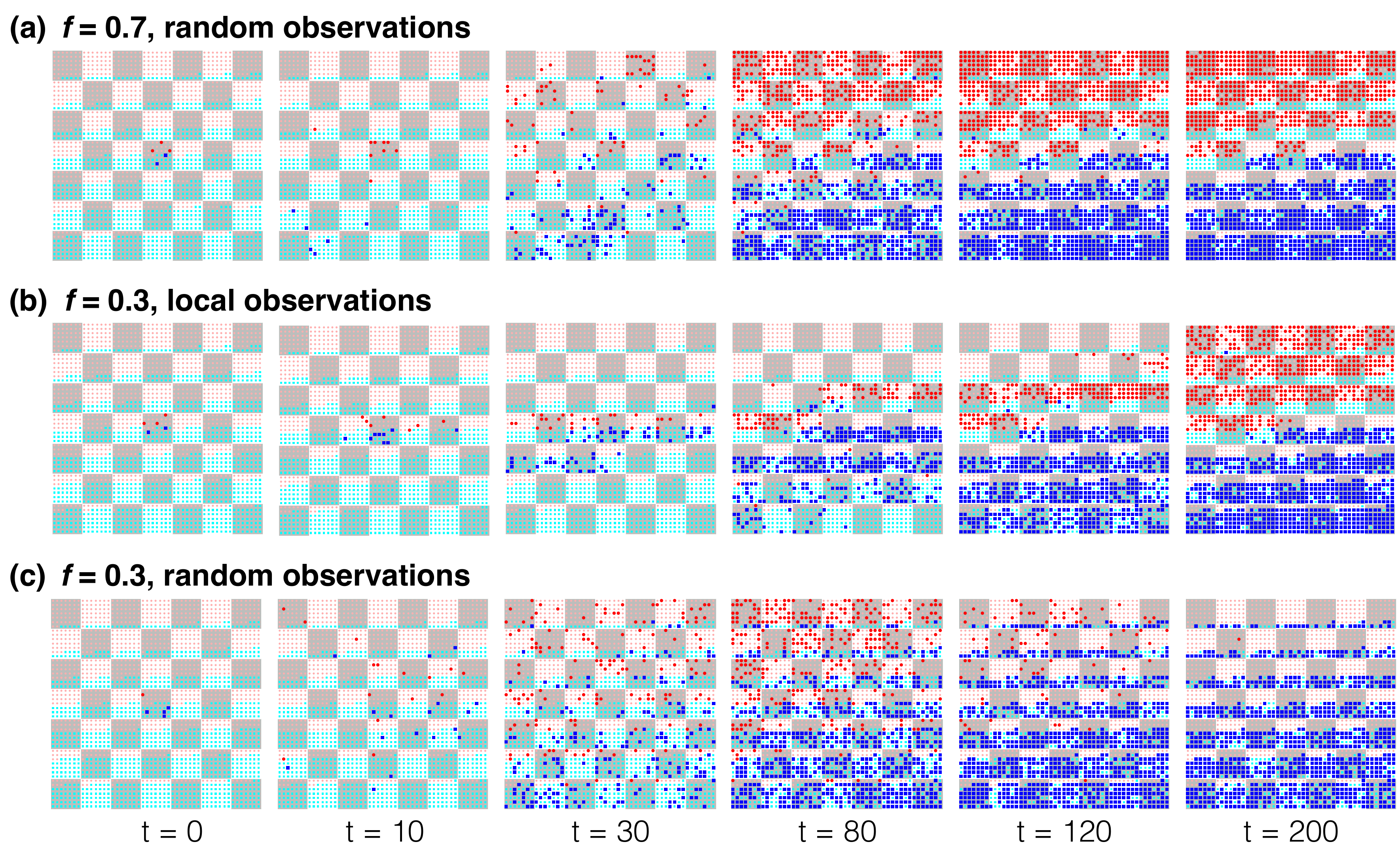}}
\caption{The dynamics of local and global polarization. Visualization of example model runs showing the spread of the innovation for different rates of and strategies for external observation. White and grey square patches each contain 49 agents. Small cyan and pink agents have not adopted. They turn dark blue and red, respectively, when they adopt. While all three runs illustrate underadoption, we also observed delayed (b) and suppressed (c) adoption.}
\label{dynscreens}
\end{figure}

\begin{figure}
\centerline{\includegraphics[width=.95\textwidth]{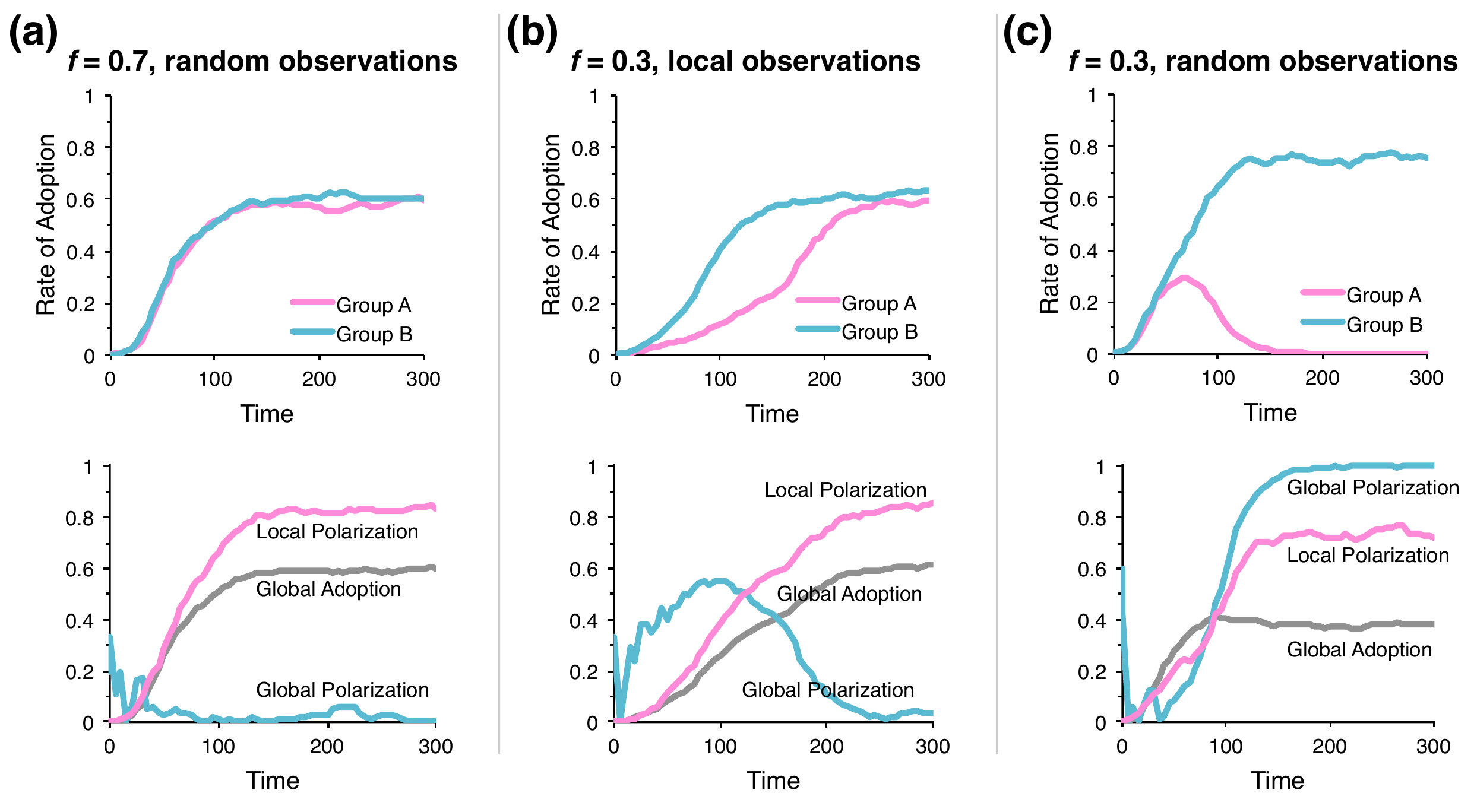}}
\caption{Dynamics of adoption and polarization, from the runs depicted in Figure \ref{dynscreens}.}
\label{dyngraphs}
\end{figure}

When observations were mostly local, we observed local but not global polarization, with persistent underadoption at the population level. Here, outgroup aversion suppressed adoption by the minority within patches (Figures \ref{dynscreens}a and  \ref{dyngraphs}a). Results here were similar regardless of whether extra-patch observations were local or random, because most observations occurred within an agent's own patch.

When external observations were more common but restricted to neighboring patches (which are demographically similar in terms of group membership), we observed delayed adoption resulting from small stochastic differences in initial conditions, which cascade as a result of network structure. Here, adoption flowed from patch to patch in the direction of increasing prevalence of the group that first began to adopt at a higher rate. In the other direction, members of the same group as those early adopters were increasingly rare, allowing outgroup aversion to delay adoption until sufficient individuals from the second group adopted. Once this occurred, adoption flowed from patch to patch in that direction among the second group
(Figures \ref{dynscreens}b and  \ref{dyngraphs}b). In such cases, the population may initially become quite polarized (as when one group is the primary adopter), but quickly equilibrates so that both groups adopted in nearly equal numbers overall. Note here that there is still underadoption at the population level, because the product spreads only among one group (typically the more numerous one) in each patch. 

When external observations were common and randomly drawn from the entire population, we observed total suppression of adoption in one group. This result emerged from small differences in initial patterns of adoption and the subsequent path dependency of stochastic interactions. When one group, purely by chance, exhibited a slightly higher frequency of  adoption early on, this initial disparity generated positive feedback, making it more likely that members of the same group would adopt and that outgroup individuals would not (Figures \ref{dynscreens}c and  \ref{dyngraphs}c).

The presence of multiple identity groups and widespread aversion toward adopting products or behaviors associated with an outgroup yield these distinct diffusion dynamics. These are the major results generated by our model. For completeness, we also present a more detailed analysis of the model behavior, and explore how the dynamics respond to variations in parameter values.

\subsection{Outgroup Aversion Hinders Adoption}

First, we consider whether outgroup aversion hinders overall levels of adoption.  For example, will the total purchases of high-end electric cars be suppressed if a large share of the population would prefer not to be associated with the West-coast elites who were among the first to drive Teslas?  Moreover, if a product is less intrinsically desirable (and thus its adoption more dependent on social influence), how will this influence patterns of adoption in our model? We expected adoption would decrease with an increase in either outgroup aversion, $b$, or frequency dependence, $\lambda$. 
These broad predictions follow from our model assumptions. However, the extent to which these expectations will be influenced by demographic skew and the location of early adopters was unclear. 

\begin{figure}
\centerline{\includegraphics[width=.95\textwidth]{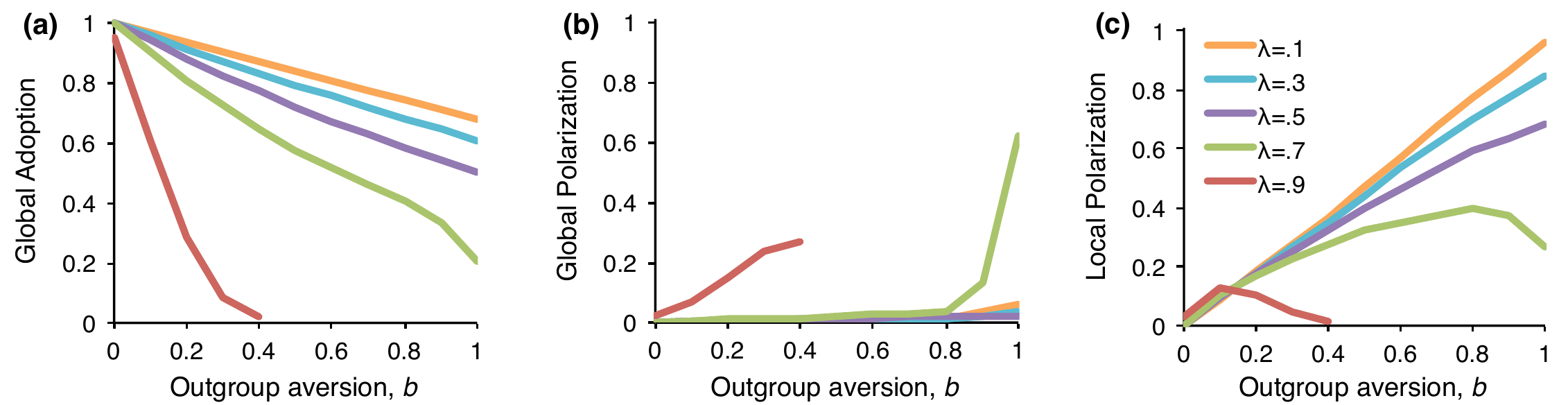}}
\caption{Effect of outgroup aversion, $b$, and frequency dependence, $\lambda$.}
\label{outgroupaversion}
\end{figure}

\begin{figure}
\centerline{\includegraphics[width=.6\textwidth]{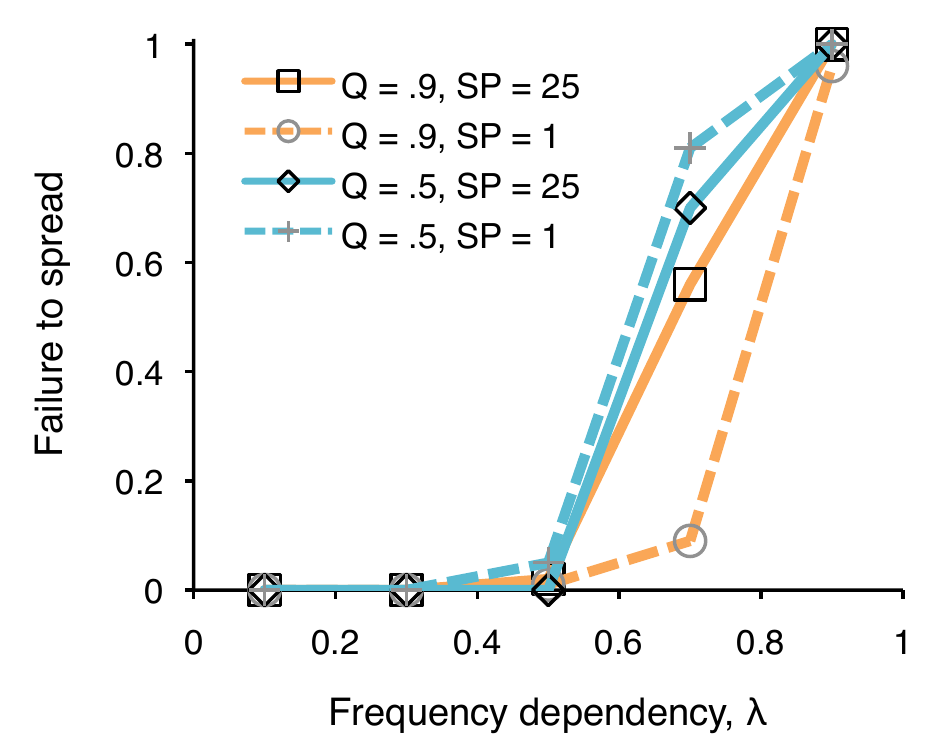}}
\caption{The fraction of runs (of 100) in which the adoption failed to spread, as a function of $\lambda$. The patch skew and starting patch were varied as indicated. }
\label{failspread}
\end{figure}

We found that increasing the strength of outgroup aversion, $b$, hindered overall adoption levels (Figure \ref{outgroupaversion}a). Stochastic events early in the adoption process led one group to become associated with the innovation within any given patch, resulting in the outgroup becoming averse to adoption. Within a patch, the adopting group was highly likely to be the local majority, both because they were more likely to be the early adopters (all things being equal) and because majority group members were more likely to be observed by potential adopters. Stronger frequency dependence also resulted in lower rates of adoption as long as outgroup aversion was nonzero. In these cases, individuals were less likely to spontaneously adopt, creating a feedback loop resulting in persistent low adoption. Indeed, because outgroup aversion reduced the availability of ready adopters and impeded the spread of the innovation when rare, high values of $\lambda$ under strong outgroup aversion sometimes meant the innovation failed entirely to spread (i.e., the adoption rate was zero at the end of the simulation), though this was also influenced by the initial location of early adopters as well as the degree of demographic skew (Figure \ref{failspread}).  The effects of high $\lambda$ were mitigated by higher demographic skew, especially when the innovation was seeded in a highly skewed patch, where effects of outgroup aversion were minimized.
 
\subsection{Outgroup Aversion Increases Local, But Not Global, Polarization}
Second, we consider the how outgroup aversion interacts with the spatial structure.  In particular, we considered whether outgroup aversion affects local and global polarization similarly. These are distinct variables; although a population may have high polarization at both local and global levels (if one group always adopts at a higher rate), it can also have high local but low global polarization (if adoption always favors the local majority). 
Our analytical model produced global polarization under some conditions, but lacked spatial structure and so could not examine local polarization. 

We found that stronger outgroup aversion did not much influence the level of global polarization (Figure \ref{outgroupaversion}b). Some global polarization did occasionally occur when frequency dependence was high, though this was due to low overall adoption rates and stochasticity in early adoption. We do note that the lack of global polarization is likely dependent on our strong assumption of equally represented social identity groups and symmetrical demographic skew. Future work should explore the effects of asymmetrically distributed social groups. That said, increased outgroup aversion reliably led to increased {\em local} polarization (Figure \ref{outgroupaversion}c). Within patches, one group tended either to have an initially higher rate of adoption (a first mover effect) or to be dominant in sheer numbers, so that the minority group of adopters came to be biased against the product. The members of the local majority group were often, but not always, the primary adopters within a patch.

\subsection{Higher Demographic Skew Increases Adoption When Interactions Are Local}
Third, we investigated how the spatial distribution of identity groups interacts with the patterns of communication between patches to influence patterns of adoption. 
Our analytical model assumed a well-mixed population, and so could not reveal any insights into this question. Analysis of our agent-based model indicates that these factors matter.  

\begin{figure}[tp]
\centerline{\includegraphics[width=.6\textwidth]{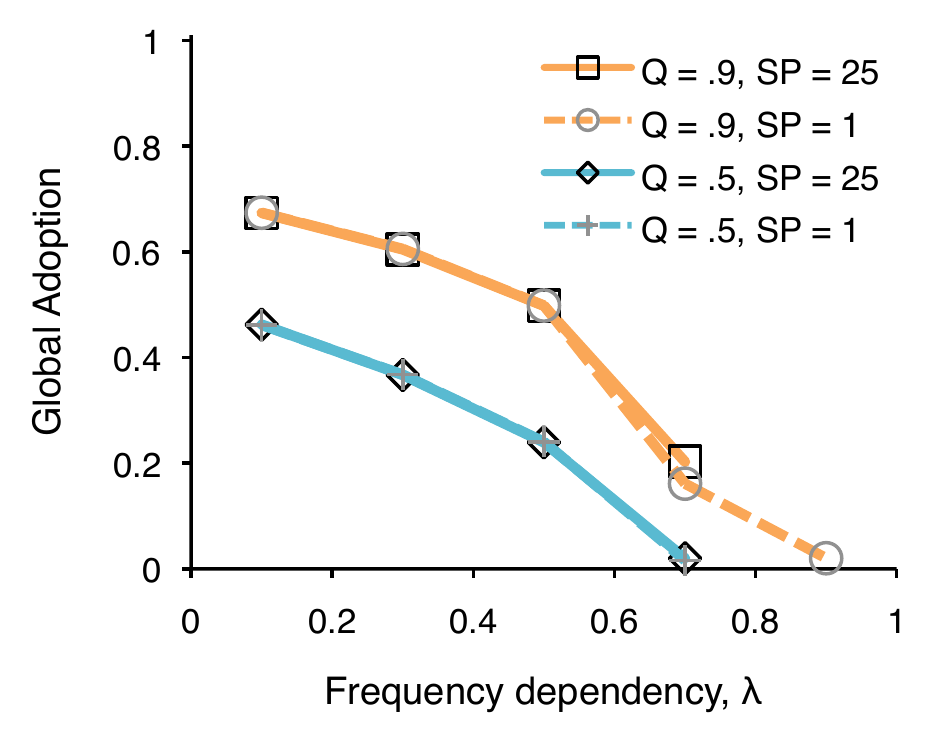}}
\caption{The global rate of adoption as a function of $\lambda$. Increased demographic skew increases rates of adoption. Adoption is unaffected by the start patch. }
\label{demskew}
\end{figure}

The demographic skew, $Q$, determines the degree to which some patches will be preferentially populated by one group over the other. $Q = 0.5$ indicates that each patch draws 50\% of its members from each group. The start patch (SP) is where the innovation is first adopted. As noted, under moderately high frequency dependence, innovations spread more readily under high demographic skew and when the location of early adopters is centrally located in the network (Figure \ref{failspread}). We also found that, regardless of the strength of frequency dependence, greater demographic skew generated higher overall rates of adoption (Figure \ref{demskew}). In the long run, adopters in any given patch tended to be members of a single identity group, which created a stable scenario in which non-adopters were prevented from ever adopting due to outgroup aversion. This influenced overall levels of adoption, and the degree of global underadoption. In a scenario in which all patches have 50\% of their members from each group, the total population-level of adoption will be at most 50\%. In contrast, a world in which patches are skewed can yield greater overall adoption, because within any patch a majority of individuals can be adopters. We note that once the innovation spread, the long-term outcome was not affected by the location of early adopters.  In other words, the location of early adopters influenced the proportion of failed runs (runs in which the innovation failed to spread) but did not have an impact on the long-term rate of overall adoption.

\subsection{The Absence of Demographic Skew Increases Global Polarization}
\begin{figure}[tp]
\centerline{\includegraphics[width=.95\textwidth]{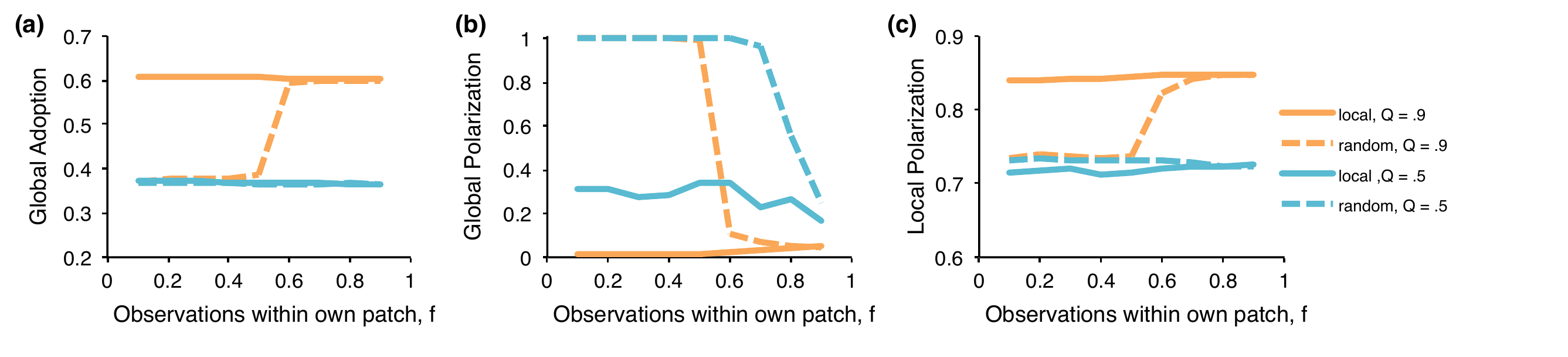}}
\caption{Aggregate statistics for runs different rates of and strategies for external observation under different levels of patch skew, Q.}
\label{randomobs}
\end{figure}

Fourth, we found that the level of demographic skew, $Q$, influenced the amount of polarization in the population, and also affected how polarization was influenced by the  rate of within-patch observation, $f$. As noted above, when patches were skewed and agents were averse to the outgroup, the population could sustain higher rates of adoption than when patches all had equal numbers of each group. When patches were skewed, the maximum adoption levels occurred as long as the rate of long-range interactions were kept relatively low, either by having a large $f$ for random observations or by restricting observations locally. However, when patches all contained an equal number of agents from either group, the overall rate of adoption was unaffected by either the rate of external observations or whether external patches are located locally or randomly (Figure \ref{randomobs}a). The adoption rate was unaffected because within each patch, one group dominated the adoption as long as some nonzero proportion of observations were made internally. The patches were therefore equally polarized (Figure \ref{randomobs}c) regardless of which group has the innovation. 

\begin{figure}[tp]
\centerline{\includegraphics[width=.95\textwidth]{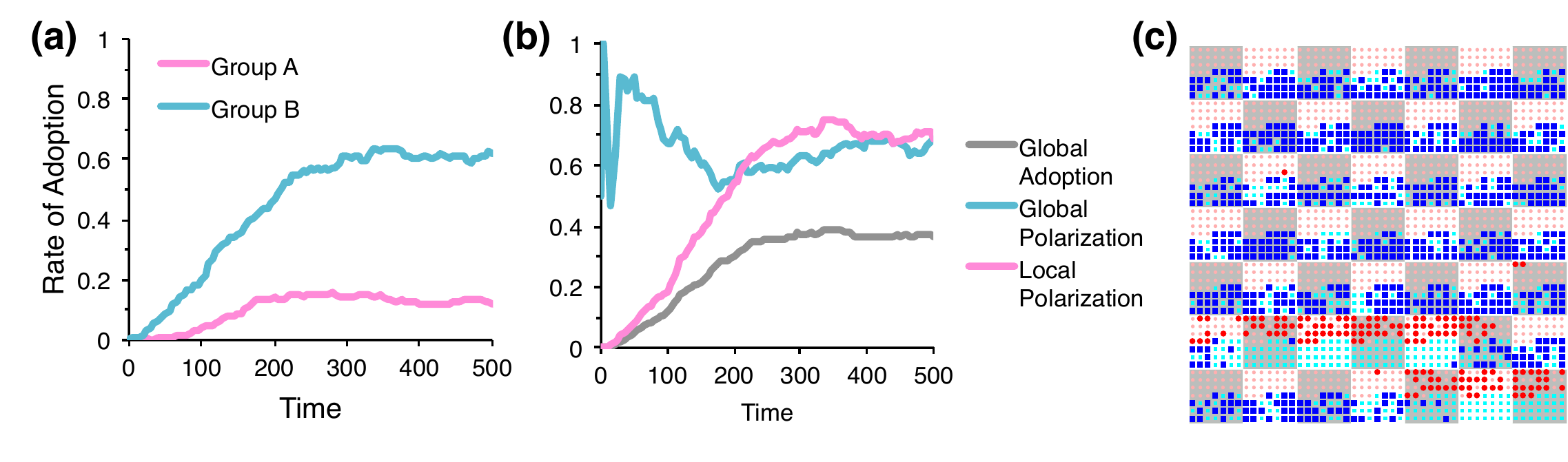}}
\caption{(a, b) Dynamics for an example run in which Q = 0.5, f = 0.3, and external observations are local. (c) Screen shot of the population at time t = 500.  }
\label{dynpol}
\end{figure}

Figure \ref{randomobs}b shows the conditions under which the population could achieve global polarization (that is, when the innovation was adopted only by one group). When patches were highly skewed, there was a transition from complete polarization to (near) zero polarization as long-range interactions become rarer. When all patches had equal numbers of each group, there was almost always some degree of polarization, even when one group did not completely dominate. Polarization continued because within each patch, the group that achieves an early majority in adoption levels is determined entirely by stochastic events. Figure \ref{dynpol} shows the dynamics of a run in which observations were local and there was no demographic skew ($Q = 0.5$).  Although the population structure did not favor either group to be the primary adopter globally, the nature of observations nevertheless facilitated significant global polarization. 

\section{Discussion}

We have constructed a model that takes population structure into account to deepen our understanding of how identity signaling, and particularly outgroup aversion, affects the dynamics of product adoption. 
We find that the emergence of products as identity signals, coupled with an aversion to be mistaken for a member of an outgroup, can dramatically influence the dynamics of how products or other cultural variants diffuse in a population structure. 
The network structure of the population and the way information is transmitted within and between social network clusters also influences the long-term dynamics of diffusion, resulting in population-wide underadoption. Under some conditions, we observe that one group may experience delayed adoption or have adoption entirely suppressed. 
We find that an aversion to adopt products associated with an outgroup can decrease overall rates of adoption, leading to local polarization within network clusters or other communities. When communication is restricted to local networks, or otherwise remains between demographically similar communities, polarization can remain local. When communication is long-range, such that interactions occur between individuals from demographically dissimilar communities, polarization can become global. Perhaps most interestingly, we find that structural constraints on information flow can generate phenomena otherwise attributable to intrinsic between-group differences.
	
Social identities are among the driving factors in organizing social behaviors in complex societies. Individuals organize into marked groups, within which norms and values are reinforced, setting differential criteria for wide ranges of social behavior, including consumer behavior. Interactions between relative strangers are commonplace in the modern industrialized world, necessitating mechanisms for partner selection and coordinated joint activities. Social identity is likely an important component of those coordination efforts \citep{smaldino2017}. Arbitrary markers, including fashion, modes of speech, or product adoption, can serve as useful signals for determining identity if they are reliable signals thereof. Importantly, such markers need not have any initial intrinsic association with an identity. Instead, when the need for coordination is present, associations between markers and meaningful identities can emerge organically \citep{dellaposta2015, efferson2008, mcelreath2003}. This association can feed back into subsequent social behavior. If the adoption of markers, including consumer products, is seen as a signal of group identity, individuals might not adopt products that they associate with outgroups \citep{berger2007, berger2008}. Our model indicates that, at the population level, this can lead to widespread polarization in the adoption of products, including delayed or suppressed adoption by one group that would otherwise find adoption appealing, as well as underadoption at the population level. 

Our results further indicate that the increases in long-range communication facilitated by technologies such as text messaging and social networking sites should lead certain products and technologies to become globally associated with certain groups or affiliations. This polarization should interact with and be reinforced by psychological forces which lead individuals to grow increasingly similar to interaction partners whose views they already share, and increasingly different from those whose views they do not share \citep{lord1979, miller1993}. Previous analysis of opinion dynamics in a network indicate that increased long-range interactions can foster increased polarization in network clusters, in part because individuals are more likely to be exposed to extreme versions of their original views \citep{flache2011}. Such polarization might create new ingroup-outgroup distinctions, leading to further reductions in the adoption of innovations. 

If more widespread adoption and the associated reduction of polarization are desirable goals, the role of social identity cannot be ignored. A pressing example is in the adoption of sustainable or environmentally ``friendly" technologies. Firms, advertisers, and policy makers should make efforts to reduce the extent to which products are seen as social markers. One possibility is increased attention to research advanced by social psychologists on how to reduce ingroup bias and turn opponents into collaborators (e.g. Sherif, 1988)\nocite{sherif1988}. Another option might be the introduction of competing but equivalent products to occupy the various niches created by different identity groups. Although the present analysis suggests that social identity salience should be avoided for maximal diffusion, this runs counter to the way many companies explicitly and successfully market their products, because in reality brands often compete for different submarkets. On the other hand, activists looking to decrease the market share of an environmentally costly product might seek to associate it with one clearly marked group to avoid adoption by another group. 

The factors influencing innovation diffusion among competing brands are many, and analysis thereof in the context of identity signaling is beyond the scope of this paper, but it is perhaps helpful to speculate as to potential complications to diffusion dynamics arising from social identity. Consider competition between brands with cross-brand influence. \cite{libai2009} modeled such a system and noted a rapid takeoff of the follower brand, which was countered by a persistent ``interaction-based" advantage to the first entrant, in which its initial numeric advantage continuously fed back to generate more new adopters. They showed these effects could be observed in cellular service markets in Western Europe in the span between 1985-2005. Identity was not necessary to explain their results, a fact we find unsurprising. We should not expect social identity to be a significant factor in this setting, because cell service is not a visible product like a car, a smartphone, or an article of clothing. For products that {\em can} serve as a social marker, we expect the dynamics of adoption to be different. Exactly how group identity and social markers influence cross-brand adoption effects is not clear, but one assumption of cross-brand influence is that the prevalence of one brand increases the likelihood of adoption for the other brand. This effect could potentially be amplified in the case of identity signaling, because there would simultaneously be a reminder of the niche being filled by the product, and an incentive to avoid identification with the outgroup. In this case, the timing of a counter-identifying brand is expected to be even more critical than previously thought. On the other hand, the timing of brand introduction may matter less if the effect of identity signaling is very strong. We performed some simple analyses with an extension of our model, not presented here, in which two competing brands were introduced simultaneously. We found that, under some conditions, each brand could become fully associated with a different identity group, following an initial period of adoption and dis-adoption as the brands emerged as markers for each group. A related possibility is that competing brands may be viewed as categorically similar, leading to cross-brand inhibition. A considered analysis of social identity in diffusion dynamics with competing brands is warranted in future research.  
	
Our model presents a picture of innovation adoption that is necessarily limited. One factor neither of our models explored is differences in population size among the two groups. We might expect that outgroup aversion, in the context of groups of different size, could actually increase the amount of adoption overall, if the larger group is the adopting group, while suppressing adoption among the minority group.  Many factors other than social identity influence decisions to adopt a product, behavior, or other cultural variant, including the innovation's intrinsic quality, the status or social power of current and potential adopters ({\em sensu} Bonacich, 1987),\nocite{bonacich1987}  and the current needs and available resources of the individual. Multiple products can exist to fit similar niches, each of which may appeal to different identity demographics. Multiple groups exist, not just two. Social identities are complex, hierarchical, and context dependent; certain group identities can become more or less salient depending on personal, social, and environmental circumstances \citep{ashmore2004, roccas2002, smaldino2017}. Moreover, identifying the network structure of communication-related product information can be tricky. Individuals are influenced by direct observation, word of mouth, advertising, and other media. Network position matters. For example, individuals with more network ties are often early adopters, while those with fewer network ties are often later adopters \citep{valente1996}. The availability of products and institutional support for their adoption may work in tandem with social identity to foster or impede support for a product. For example, identification as a political liberal or environmentalist might lead a person to become interested in adopting hybrid or electric vehicle technologies, but the feasibility of that adoption is highly dependent on the presence of local dealerships and the infrastructure of charging stations, which in turn may depend on government incentives \citep{diamond2009, sanroman2011, wirasingha2008}. Nevertheless, the simplifications in our model allow an uncluttered examination of the role of social identity and emergent social markers in the dynamics of innovation adoption. 

\section*{Acknowledgments}
\small
The authors acknowledge the support of the National Institute for Mathematical and Biological Synthesis (NIMBioS), an Institute sponsored by the National Science Foundation through NSF Award \#DBI-1300426, with additional support from The University of Tennessee, Knoxville. We thank the members of NIMBioS working group, Evolutionary Approaches to Sustainability, for valuable discussion. For comments on an earlier version of this manuscript, we thank Jeremy Brooks, Fred Feinberg, Cristina Moya, Karthik Panchanathan, and Tim Waring.  We are particularly grateful to Tim Waring for leadership in the NIMBioS working group.

\bibliographystyle{apacite} 
\bibliography{socialadoption.bib}

\end{document}